\documentstyle[psfig]{acm}

\title{GENERAL PRINCIPLES OF LEARNING-BASED MULTI-AGENT SYSTEMS}

\author{David H. Wolpert\\
NASA Ames Research Center\\
Moffett Field, CA 94035\\
dhw@ptolemy.arc.nasa.gov\\
\and
Kevin R. Wheeler\\
NASA Ames Research Center\\
Caelum Research \\
Moffett Field, CA 94035\\
kwheeler@mail.arc.nasa.gov\\
\and
Kagan Tumer\\
NASA Ames Research Center\\
Caelum Research \\
Moffett Field, CA 94035\\
kagan@ptolemy.arc.nasa.gov }

\begin{document}

\maketitle

\begin{abstract}
We consider the problem of how to design large decentralized multi-agent systems
(MAS's) in an automated fashion, with little or no hand-tuning. Our
approach has each agent run a reinforcement learning
algorithm. This converts the problem into one of how to automatically
set/update the reward functions for each of the agents so that the
global goal is achieved. In particular we do not want the agents to
``work at cross-purposes'' as far as the global goal is concerned.
We use the term artificial COllective INtelligence
(COIN) to refer to systems that embody solutions to this problem.
In this paper we present a summary of a mathematical
framework for COINs. We then investigate the real-world applicability
of the core concepts of that framework via two computer experiments:
we show that our COINs perform near optimally in a difficult variant
of Arthur's bar problem~\cite{arth94} (and in particular avoid the
tragedy of the commons for that problem), and we also illustrate
optimal performance for our COINs in the leader-follower problem.
\end{abstract}

\section{INTRODUCTION}
\label{sec:intro}
In this paper we are interested in computational problems having the
following characteristics: 
 
\noindent $\bullet$  the agents each run reinforcement learning (RL) algorithms; 

\noindent  $\bullet$ there is little to no centralized communication or control;

\noindent  $\bullet$ there is a provided world utility function that rates the
possible histories of the full system.

There are many examples of such problems, some of the more prominent
being:

\begin{enumerate}
\item Designing a control system for constellations of communication
satellites or of constellations of planetary exploration vehicles
(world utility in the latter case being some measure of quality of
scientific data collected);
\item Designing a control system for routing over a communication
network (world utility being some aggregate quality of service
measure);
\item Construction of parallel algorithms for solving numerical
optimization problems (the optimization problem itself constituting
the world utility);
\item Control of a large, distributed chemical plant.
\end{enumerate}

These kinds of problems may well be most readily addressed by using a
large Multi-Agent System (MAS)~\cite{sen97}, where each agent is
restricted to communicate with only a few neighbors, and where each
agent runs a Reinforcement Learning (RL) algorithm.  In such systems,
a crucial problem is ensuring that the agents' RL algorithms do not
``work at cross-purposes'', so that their collective behavior
maximizes a provided global utility function.  The difficulty in
achieving this is that these systems have no centralized control, so
the dynamics is governed by the collective effects of the individual
agents each modifying their behavior via their (local) RL algorithms.

We are interested in such systems where
the agents are ``greedy'' (i.e., there is no external structure
forcing cooperation). 
However, both the agents reward functions and the overall system structure are 
automatically set and then updated in a machine learning-like fashion, 
so as to facilitate the achievement of the global objective.

As opposed to hand-tailored MAS design, these alternative approaches
potentially have the following benefits: one does not have to
laboriously model the entire system; global performance is ``robust'';
one can scale up to very large systems; and one can maximally exploit
the power of machine learning.  We use the term COllective
INtelligence (COIN)~\cite{wotu99b,wotu99a,wowh99b} to refer to either MAS's
designed in this way, or (in the case of naturally occurring MAS's) to
MAS's investigated from this perspective.

The COIN framework is related to many other fields.  (See
~\cite{wotu99b} for a detailed discussion of these relations, involving
several hundred references.) Some of them are:
\begin{itemize}
\item multi-agent systems;~\nocite{sen97}  
\item computational economics;~\nocite{bosh97,cama97} 
\item reinforcement learning for adaptive control;~\nocite{crba96}
\item statistical mechanics;~\nocite{been98}
\item computational ecologies;~\nocite{huho88}
\item game theory~\nocite{sala98}, in particular, evolutionary game theory.~\nocite{chzh97} 
\end{itemize}
Previous MAS's most similar to a COIN include
those where agents use reinforcement learning~\cite{clbo98,huwe98b},
and/or where agents actively attempt to model the behavior of other
agents~\cite{huwe98a}.

In this paper we introduce some of the concepts from the COIN
framework, and then present experiments testing those concepts.  The
restricted version of the framework presented here is not sufficient
to formally guarantee optimal global performance for all multi-agent
systems.  Rather the experiments recounted below are designed to
empirically investigate the usefulness of these concepts in some
illustrative domains.  In Section~\ref{sec:mot} we present general
background on COINs and the experiments we conducted.  In
Section~\ref{sec:math}, we present the portion of the COIN framework
investigated in this paper.  In Sections~\ref{sec:bar} and
\ref{sec:leader}, we describe the bar and leader-follower experiments,
respectively, and how our approach deals with the many pitfalls
encountered in each.

\section{MOTIVATION AND BACKGROUND}
\label{sec:mot}

A naturally occurring example of a system that can be viewed as COIN
is a human economy.  For example, one can take the agents to be the
individuals trying to maximize their personal rewards. One might then
declare that the world utility is a time average of the gross domestic
product. (``World utility'' per se is not a construction internal to a
human economy, but rather something defined from the outside.) To
achieve high global utility it is necessary to avoid having the agents
work at cross-purposes lest phenomena like the Tragedy of the Commons
(TOC) occur, in which individual avarice works to lower global
utility~\cite{hard68}. One way to avoid such phenomena is by modifying
the agents' utility functions via punitive legislation.  A real world
example of an attempt to make such a modification was the cration of
anti-trust regulations designed to prevent monopolistic practices.

In designing a COIN we have more freedom than anti-trust regulators
though, in that there is no base-line ``organic'' local utility
function over which we must superimpose legislation-like
incentives. Rather, the entire ``psychology'' of the individual agents
is at our disposal when designing a COIN.  This freedom is a major
strength of the COIN approach, in that it obviates the need for
honesty-elicitation mechanisms, like auctions, which form a central
component of conventional economics.

We recently investigated the use of the COIN approach for distributed
control of network packet routing~\cite{wotu99a}. 
Conventional approaches to
packet routing have each router run
a shortest path algorithm (SPA), 
i.e., each router routes its packets in the way that it expects will
get those packets to their destinations most quickly. 
Unlike with a
COIN, with SPA-based routing the routers have no concern for the
possible deleterious side-effects of their routing decisions on the
global goal (e.g., they have no concern for whether they induce
bottlenecks). 
We ran simulations that demonstrated that a COIN-based routing 
system has better throughputs than does an SPA-based system~\cite{wotu99a}.

In this paper we present a more fine-grained investigation of our COIN
methodology, in which we disentangle two of the major components of
that methodology and examine the real-world usefulness of them
separately. 
In the first set of experiments we investigate how one might initialize 
a COIN (i.e., initialize each agents' local reward function) to ensure that the 
agents do not work at cross purposes.
The problem we chose for this purpose is
a more challenging variant of
Arthur's bar attendance problem~\cite{arth94,chzh97,joja98}. In this problem,
agents have to determine which night in the week to attend a bar.
The problem is set up so that if either
too few people attend (boring evening) or too many people attend (crowded
evening), the total enjoyment of the attendees drops.
Our goal is to design the reward functions of the attendees 
so that the total enjoyment across all nights is maximized.
(This problem is similar to a number of problems arising in 
electronic commerce, where agents 
need to select the ``best'' markets to trade their wares.)

In the second set of experiments we investigate how run-time 
modification of agent utility functions and/or inter-agent interactions
can improve performance beyond that of the initialized COIN.
For this set of experiments we chose a problem where
certain ``follower'' agents mimic the activities of other 
``leader'' agents.
The idea is to investigate a scenario where 
actions that are beneficial to a particular 
agent have deleterious global effects when copied 
by other agents.

\nocite{brad97,futi91,fule98,jesy98,syca98,baum98}

\section{COIN FRAMEWORK}
\label{sec:math}
As applied here, the COIN framework has three main components: 

\begin{enumerate}
\item
The first component investigates various formalizations of
the desideratum that whenever the local utility functions increase,
then so must the value of the world utility. It implicitly defines
equivalence classes of local utility functions that meet those
formalizations, for allowed agent interactions of various types. 
\item
There is little explicit concern for the dynamical nature of the COIN
in this first component of the framework; that dynamics is subsumed under
the assumption that the agents can achieve large values of their local
utility functions, via their RL learning. However due to that
dynamics it may be that the agents are all inadvertently
``frustrating'' each other, so that none of them can achieve large
values of their local utility functions. The second component of the
COIN framework addresses this problem by means of further restrictions 
(beyond those of the first component) on the allowed set of local 
utility functions. 
We call the design of a system's local utility functions using
the two components of the COIN framework the {\it COIN initialization} 
of that system.
\item
Successful COIN initialization ensures that if the agents in the
system achieve large values of their local utility functions, then the
world utility is also large. However due to uncertainties in the type
of the agent interactions, in the real world the local utility
functions set in the COIN initialization are usually only a good first
guess. 
The third component of the COIN framework addresses this issue
by modifying the local utility functions at
run-time based on localized statistical information. 
We call this modifying process {\it macrolearning}.
In contrast, we call the reinforcement learning that agents perform to
optimize their local reward {\it microlearning}. 
\end{enumerate}

In this paper we 
consider the state of the system across a set of discrete,
consecutive time steps, $t \in \{0, 1, ...\}$. 
Without loss of generality, we let all relevant
characteristics of an agent at time $t$ --- including its internal
parameters at that time as well as its externally visible actions ---
be encapsulated by a Euclidean vector $\underline{\zeta}_{\eta,t}$. We
call this the ``state'' of agent $\eta$ at time $t$, and let 
$\underline{\zeta}$ be the state of all agents across all time. 
World utility, $G(\underline{\zeta})$, is a function of the 
state of all agents across all time. Our goal as COIN designers is to
maximize world utility.

The following concepts and definitions form the core of the COIN framework: \\

\noindent{\bf Subworlds:}
Subworlds are the sets making up an exhaustive
partition of all agents. For each subworld, $\omega$, all agents in that
subworld have the same {\it subworld utility function}
$g_{\omega}(\underline{\zeta})$ as their
local utility functions. Accordingly, consider having each subworld be
a set of agents that collectively have the most effect on each other.
In this situation, by and large, agents cannot work
at cross-purposes, since all agents that affect each other 
substantially share the same local utility. \\

\noindent{\bf Constraint-alignment:}
Associated with subworlds is the concept of a (perfectly) {\it
constraint-aligned} system. That is a system in which any
change to the state of the agents in subworld $\omega$ at time 0 will
have no effect on the states of agents outside of $\omega$ at times
later than 0.
Intuitively, a system is constraint-aligned if no two agents in separate
subworlds affect each other, so that the rationale
behind the use of subworlds holds. (In the real world of course,
systems are rarely exactly constraint-aligned.) \\

\noindent{\bf Subworld-factored:}
A {\it subworld-factored} system is one where
for each subworld $\omega$ considered by itself, a change at time
0 to the states of the agents in that subworld results in an increased
value for $g_{\omega}(\underline{\zeta})$ if and only if it results in an
increased value for $G(\underline{\zeta})$.
For a subworld-factored system, the side effects on the rest of the system
of $\omega$'s increasing its own utility
do not end up decreasing
world utility. For these systems, the separate agents successfully
pursuing their separate goals do not frustrate each other
as far as world utility is concerned.

The definition of subworld-factored is carefully crafted. In
particular, it does {\it not} concern changes in the value of the
utility of subworlds other than the one changing its state. It also
does not concern changes to the states of agents in more than one
subworld at once.  Indeed consider the following property: any change
at time 0 to the entire system that improves all subworld utilities
simultaneously also improves world utility. This might seem an
appealing alternative desideratum to subworld-factoredness. However
one can construct examples of systems that obey this property and yet
quickly evolve to a $minimum$ of world utility (for example this is
the case in the TOC). 

It can be proven that for a subworld-factored system, 
when each of the agents' reinforcement learning algorithms are performing 
as well as they can, given each others' behavior, world utility
is at a critical point.
Correct global behavior corresponds to learners reaching a Nash
equilibrium~\cite{wotu99b}.  There can be no tragedy of the commons for a
subworld-factored system. \\

\noindent{\bf Wonderful life utility:}
Let $\mbox{CL}_{\omega}(\underline{\zeta})$ be defined as the vector
$\underline{\zeta}$ modified by clamping the states of all agents in
subworld $\omega$, across all time, to an arbitrary fixed value, here
taken to be 0. The {\it wonderful life
subworld utility} (WL) is:
\begin{equation}
g_{\omega}(\underline{\zeta}) \equiv
G(\underline{\zeta}) - G(\mbox{CL}_{\omega}(\underline{\zeta})) \; .
\end{equation}

When the system is constraint-aligned, so that, loosely speaking,
subworld $\omega$'s ``absence'' would not affect the rest of the
system, we can view the WL utility as analogous to the change in world
utility that would have arisen if subworld $\omega$ ``had never
existed''.
(Hence the name of this utility - cf. the Frank Capra movie.) 
Note however, that $\mbox{CL}$ is a purely mathematical operation.
Indeed, no assumption is even
being made that $\mbox{CL}_{\omega}(\underline{\zeta})$ is consistent
with the dynamics of the system. The succession of
states the agents in $\omega$ are clamped to in the definition of the
WL utility need not obey the dynamical laws of the system.

This dynamics-independence is a crucial strength of the WL utility. 
It means that to evaluate the WL utility we do {\it not} try to
infer how the system would have evolved if all agents in $\omega$ were
set to 0 at time 0 and the system evolved from there. So long as we
know $\underline{\zeta}$ extending over all time, and so long as we
know $G$, we know the value of WL utility. This is true even if we
know nothing of the dynamics of the system.

Another crucial advantage of the WL utility arises from the fact that
in a COIN, each agent, by itself, is operating in a large system of
other agents, and therefore may experience difficulty discerning the
effects of its actions on its utility. Very often this problem is
obviated by using the WL utility; the subtraction of
the clamped term in the WL utility removes some of
the ``noise'' of the activity of agents in other subworlds, 
leaving only the underlying
``signal'' of how the agents in the subworld at hand affect the utility. 
This makes it easier for the microlearning to improve the utility of 
the agents in that subworld. 
When the system is subworld factored, this property 
facilitates the improvement of world utility.
In many circumstances, this cancelling characteristic of WL also 
ensures that the utility functions obey the applicable {\it a priori} locality
restrictions on communications among agents.

\vspace{5mm}

The experiments in this paper revolve around the following fact: a
constraint-aligned system with wonderful life subworld utilities is
subworld-factored.  Combining this with our previous result that
subworld-factored systems are at equilibrium at critical points of
world utility, this result leads us to expect that a
constraint-aligned system using WL utilities in the microlearning will
approach near-optimal values of the world utility.

No such assurances accrue to WL utilities if the system is not
constraint-aligned however. Accordingly our first set of experiments
investigates how well a particular system performs when WL utilities
are used but little attention is paid to ensuring that the system is
constraint-aligned. Our second set of experiments then focus on a
macrolearning algorithm that modifies subworld memberships
dynamically, so as to increase the degree of constraint-alignment, all
while using WL local utility functions.

\section{BAR PROBLEM}
\label{sec:bar}
\subsection{Experimental Method}
We conducted two sets of experiments to investigate the COIN
framework. The first set 
is a variant of Brian Arthur's bar-attendance model~\cite{arth94}. 
Since we are not interested here in directly comparing our results to
those in~\cite{arth94,huwe98b}, we use a more conventional (and
arguably ``dumber'') reinforcement learning algorithm than the ones 
investigated in~\cite{arth94,huwe98b}.  

There are $N$ agents, each of whom picks one of seven nights to attend
a bar the following week, a process that is then repeated.  In each
week, each agent uses its own reinforcement learning (RL) algorithm to
decide which night to attend to maximize its utility.  

The world utility is of the form:
\begin{eqnarray*}
G(\underline{\zeta})~=~ \sum_{t} \sum_{k=1}^7 \gamma_k(x_k(\underline{\zeta},t))
\end{eqnarray*}
where ``$x_j(\underline{\zeta}, t)$'' means the $j$'th component 
of $x(\underline{\zeta}, t)$ which is the attendance on night $j$ at week $t$; 
 $\gamma_k(y) ~\equiv~ \alpha_k y \exp{(-y/c)}$; 
$c$ and each of the \{$\alpha_k$\} are real-valued parameters.
Intuitively, this world utility is the sum of the world
``rewards'' for each night in each week.

This $G$ is chosen to reflect the effects in the bar 
as the attendance profile of agents changes.  In the case when there are
too few agents, the bar suffers from lack of activity and 
therefore the world utility should be low.  
Conversely, when there are too many agents the bar becomes overcrowded and 
the rewards should also be low.  Note that $\gamma_k(\cdot)$ reaches its
maximum when its argument equals $c$.

\begin{figure*}[htb]
   \centerline{\mbox{
   	\psfig{figure={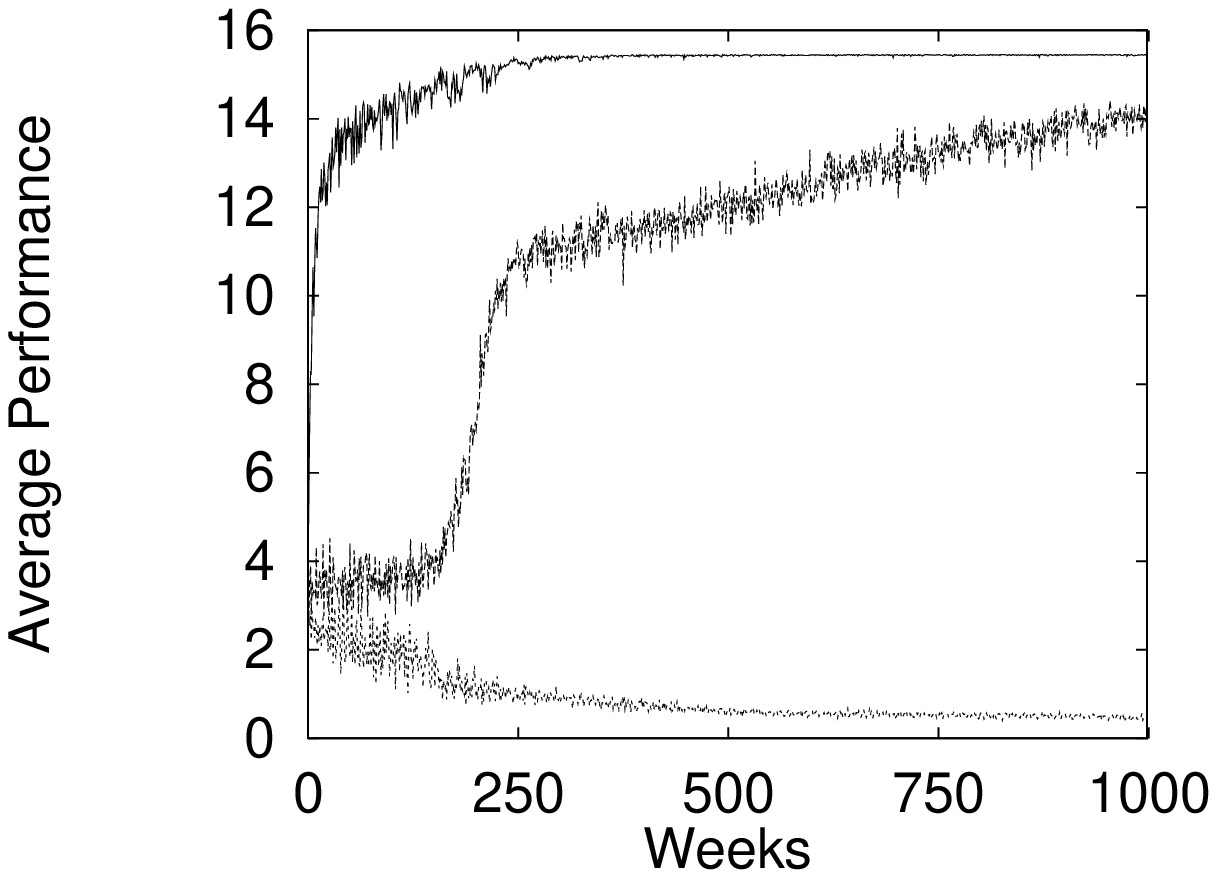},width=3.00in,height=2.4in}
	\hspace{0.10in}
   	\psfig{figure={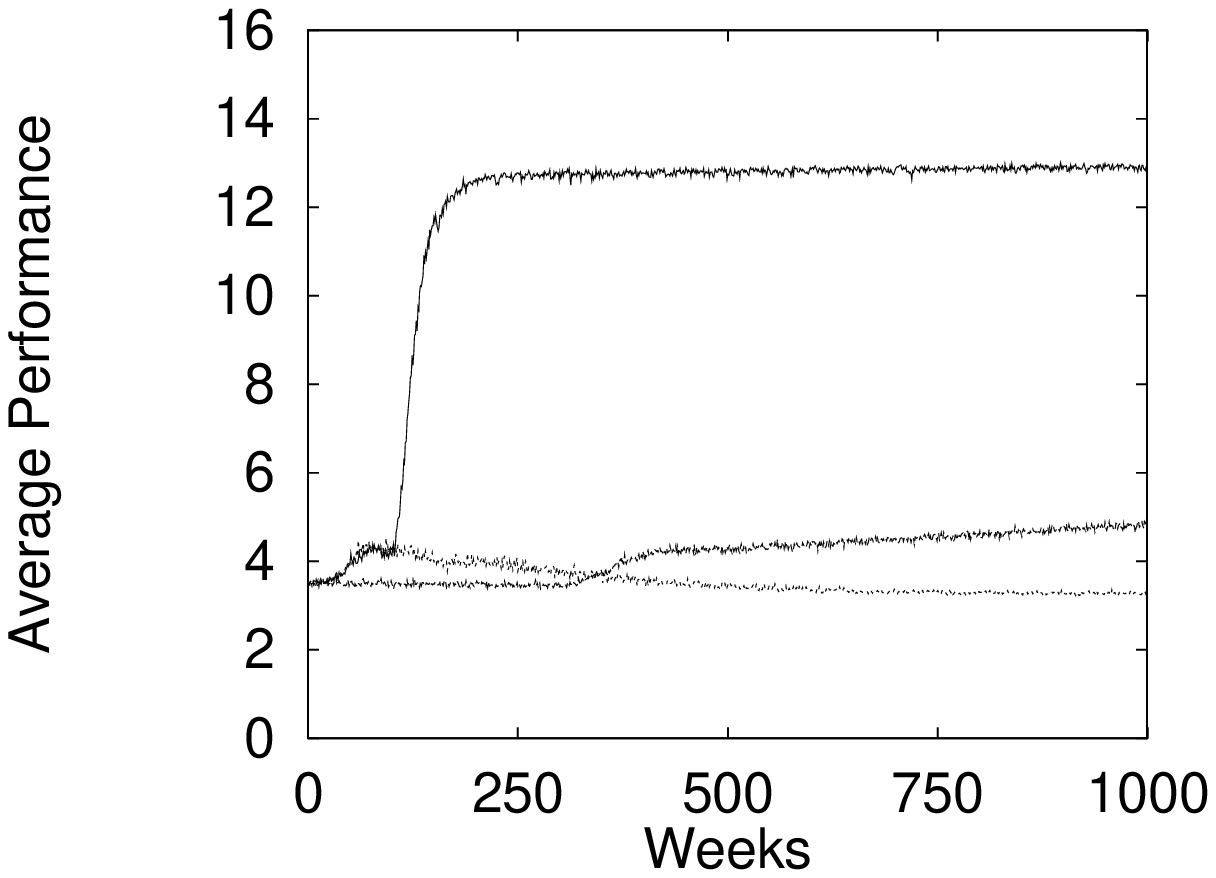},width=3.00in,height=2.4in}
   }}
  \caption{Average performance when $\alpha~=~[0~0~0~7~0~0~0]$ (left) and
	when $\alpha~=~[1~1~1~1~1~1~1]$ (right). In both the top curve is
	WL, middle is GR, and bottom is UD.}
\label{fig:barfig}
\end{figure*}

The $\alpha$'s are used to weight the importance of different nights.
Two different choices of the \{$\alpha_k$\} are investigated.  One
treated attendance on all of the nights equally, $\alpha~=~[1~1~1~1~1~1~1]$.  
The other is only concerned with the
attendance on one night of the week, 
$\alpha~=~[0~0~0~7~0~0~0]$. 
In our experiments, $c$ was 6 and $N$ is
chosen to be significantly larger (4 times) than the number of
agents necessary to have $c$ agents attend the bar on each of the 
seven nights, resulting in 168 agents.

As explicated below, our microlearning algorithms worked by providing
a real-valued ``reward'' signal to each agent at each week $t$. Each
agent's reward function is a surrogate for an associated utility
function for that agent. The difference between the two functions is
that the reward function only reflects the state of the system at one
moment in time (and therefore is potentially observable), whereas the utility
function reflects the agent's ultimate goal, and therefore can depend
on the full history of that agent across time.

We investigated the following functions specifying the rewards at time $t$
for each of the agents:
\begin{itemize}
\item Uniform Division Reward: 
\begin{eqnarray*}
   UD(d_{\omega}(t), \underline{\zeta}, t) \; \equiv \;  
   \gamma_{d_\omega}(x_{d_{\omega}}(\underline{\zeta},t))/x_{d_{\omega}}(\underline{\zeta}, t)
\end{eqnarray*}
\item Global Reward:
\begin{eqnarray*}
   GR(d_{\omega}(t), \underline{\zeta}, t) \;  \equiv \;\;
    \sum_{k=1}^7 \gamma_k(x_k(\underline{\zeta},t)) \;\;\;\;\;\;\;\;\;\;  
\end{eqnarray*}
\item Wonderful Life Reward: 
\begin{eqnarray*}
  &\!\!\!\!\!\!\!WL(&\!\!\!\!\!d_{\omega}(t),\underline{\zeta}, t) 
  	\;  \equiv \;\;\; \\
  & &   \sum_{k=1}^7  \gamma_k(x_k(\underline{\zeta},t)) - 
   \sum_{k=1}^7 \gamma_k(x_k(\mbox{CL}_{\omega}(\underline{\zeta}),t))  =  \\
  & &  \gamma_{d_\omega}(x_{d_\omega}(\underline{\zeta},t)) - 
   \gamma_{d_\omega}(x_{d_\omega}(\mbox{CL}_{\omega}(\underline{\zeta}),t))
\end{eqnarray*}
\end{itemize}
where $d_\omega$ is the night selected by subworld $\omega$. Note the
distinction between utilities and rewards. For example, world utility
is the sum over all time of global reward.

The UD reward is considered a natural ``naive'' choice of an agent's
reward function; the total reward on each night gets uniformly divided
among the agents attending that night.

Providing the GR reward at time $t$ to each agent is considered a
reasonable way to provide that agent an approximation to the value of
world utility that would ensue if the current policies of all the
agents were henceforth frozen to their current polices.  This reward
function results in all agents receiving the same feedback
information.  For this reward, the system is automatically
subworld-factored.  However, evaluation of this function requires
centralized communication.

Similarly, the WL reward at time $t$ is considered a reasonable 
approximation to the WL utility. 
However, in contrast to the GR reward, 
to evaluate its WL reward each agent only
needs to know the total attendance on the night it attended.
In addition, as
indicated above, one would expect that with the WL reward the microlearners
can readily discern the effects of their rewards on their utilities 
--- something not necessarily true if they use the GR reward.

Each agent is its own subworld.  Each agent's microlearning
algorithm uses a seven dimensional Euclidean vector to represent the
expected reward for each night of the week.  At the end of each week,
the component of this vector corresponding to the night attended is
proportionally adjusted towards the actual subworld reward just
received.  At the beginning of each week, the agent picks the night to
attend using a Boltzmann distribution with energies given by the
components of the vector of expected rewards.  A decaying temperature
is used to aid exploration in the early stage of the process.  The
microlearning algorithms are set to use the same parameters
(i.e. learning rate, Boltzmann temperature, decay rates) for all three
reward functions.  This learning algorithm is equivalent to Claus and
Boutilier's~\cite{clbo98} independent learner algorithm for
multi-agent reinforcement learning.

\subsection{Results}
Figure~\ref{fig:barfig} presents world reward values as a 
function of time for the bar problem, averaged over 50 separate runs. 
Note that world utility is the sum over time of world reward. 
We present the performance for all three 
subworld reward functions for both $\alpha~=~[1~1~1~1~1~1~1]$ and 
$\alpha~=~[0~0~0~7~0~0~0]$. 

Systems using the WL reward converged to optimal performance.  
This indicates that the bar 
problem is sufficiently constraint-aligned so that interactions between
subworlds does not diminish performance. 
Presumably this is
true because the only interactions between subworlds occurred
indirectly, via the microlearning. 

Moreover, since the WL reward is easier for the microlearners to exploit 
(see above), one would expect the convergence using the WL
reward to be far quicker than that using the GR reward.  The GR reward
does eventually converge to the global optimum.  This is in agreement
with the results obtained by Crites~\cite{crba96} for the bank of elevators
control problem.  However, when $\alpha~=~[0~0~0~7~0~0~0]$ the GR 
reward converged in 1250 weeks.  This is more than 4 times the convergence
time for the WL reward.  When $\alpha~=~[1~1~1~1~1~1~1]$ the 
GR reward took 6500 weeks to converge, which was more than {\it 30 
times} the time WL reward took to converge.  This slow convergence is 
a result of the reward signal being diluted by the large number 
of agents in the system.

In contrast to the behavior for reward functions based on the COIN
framework, use of the UD subworld reward function results in very
poor world reward values that deteriorated as the microlearning
progressed. For the case where world reward only depends on a single
night, this is essentially an instance of the tragedy of the commons; it
is in every agent's interest to attend the same night, and their doing
this shrinks the world reward ``pie'' that must be divided among all
agents.

Taken together, these experiments demonstrate that the initialization
prescriptions of the COIN framework can result in excellent global
performance.

\section{LEADER-FOLLOWER PROBLEM}
\label{sec:leader}
\subsection{Experimental Method}
In the experiments recounted in the previous section the system 
was sufficiently constraint-aligned for the WL reward to result
in optimal performance.  
The second set of experiments involving
leaders and followers also used WL reward. 
These experiments investigated the use of macrolearning
to make an initially non-aligned system more constraint-aligned,
and thereby improve the world utility.
In this experiment, macrolearning consists of changing the 
subworld memberships of the agents.  

\begin{figure*}[ht]
   \centerline{\mbox{
      \psfig{figure={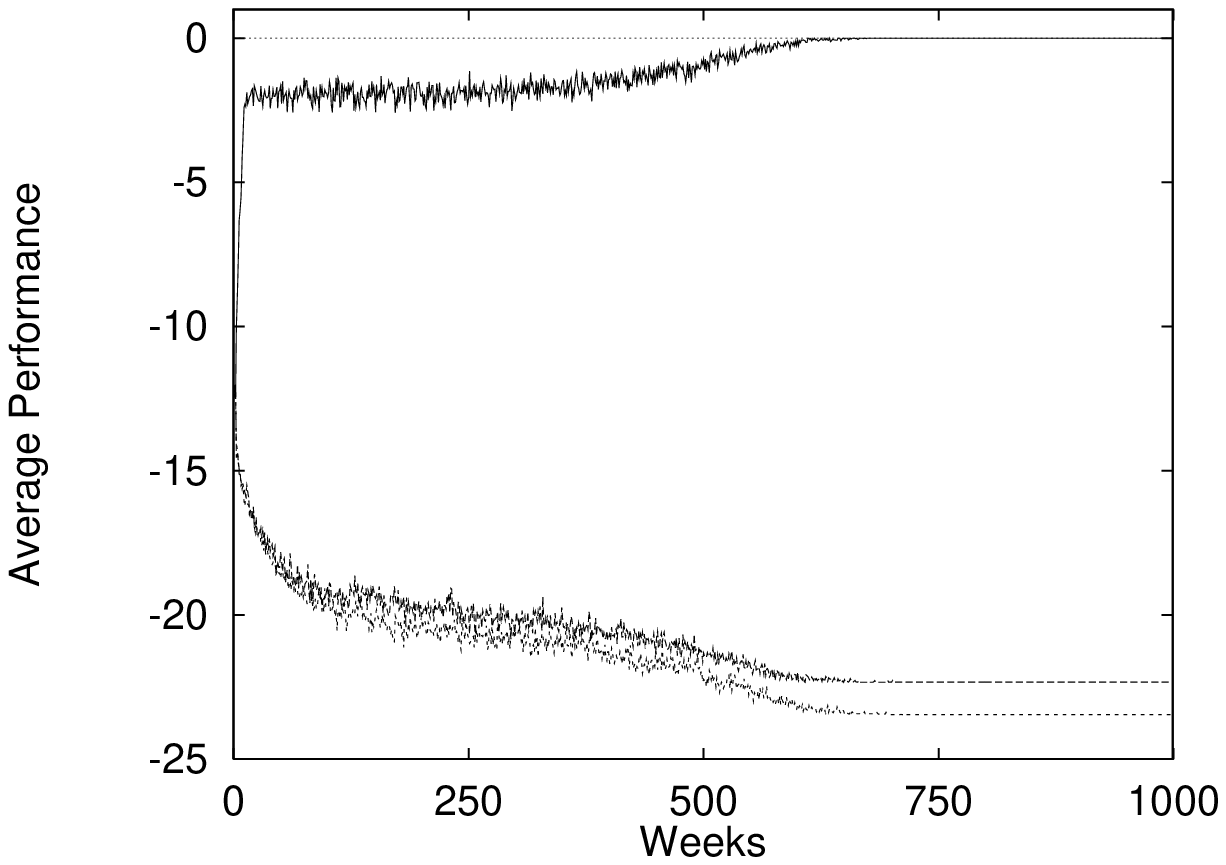},width=3.0in,height=2.25in}
      \hspace{0.1in}
      \psfig{figure={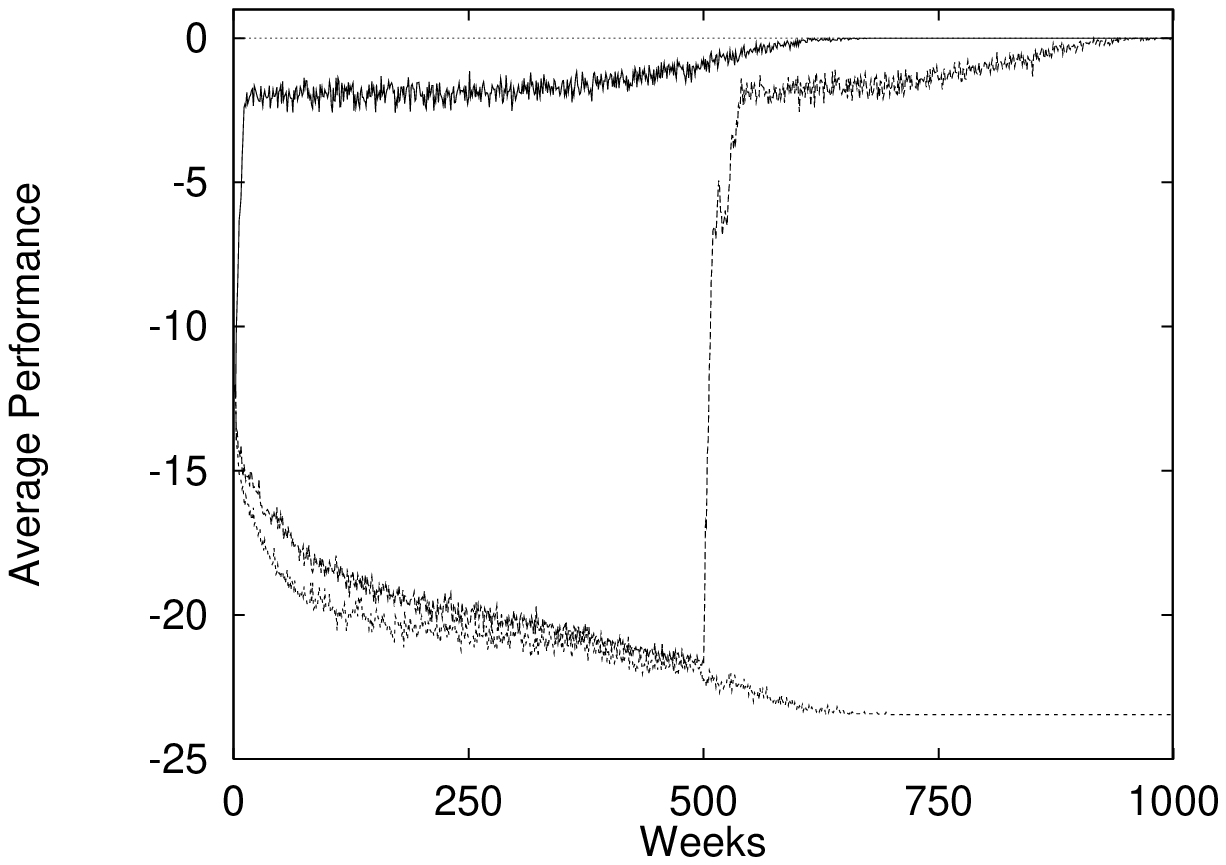},width=3.0in,height=2.25in}
   }}
   \caption{Leader-follower problem with worst case reward matrix. 
   In both plots the
   top curve represents perfect constraint-alignment, the bottom
   curve represents minimal constraint-alignment, and the middle
   curves represent random subworlds without (left)  and with (right) 
   macrolearning at 500 weeks.} 
   \label{fig:worstreward}
\end{figure*}

In these experiments the bar problem
is modified to incorporate constraints designed to frustrate
WL subworld reward.  This is done by forcing the nights picked by
some agents (followers) to agree with those picked by other
agents (leaders). Each leader has two followers.  The
world utility is the sum, over all leaders, of the values of a
triply-indexed reward matrix whose indices are the the nights that the
leader and his two followers attend: 
\begin{eqnarray*}
G(\underline{\zeta}) ~=~ \sum_t \sum_i
R_{l_i(t),f1_{i}(t),f2_{i}(t)} 
\end{eqnarray*}
where $l_i(t)$ is the night the $i^{th}$ leader
attends, and $f1_{i}(t)$ and $f2_i(t)$ are the nights attended by the
followers of leader $i$, in week $t$.  
The system's dynamics is
what restricts all the members of each triple $(l_i(t), f1_i(t),
f2_i(t))$ to equal the night picked by leader $i$ for week $t$. 
However, $G$ and $R$ are defined for all possible triples, $l_i(t), f1_i(t)$ and $f2_i(t)$. 

So in particular, $R$ is defined for dynamically unrealizable triples
that arise in the clamping operation.
Because of this, for certain $R$'s there exists subworld memberships 
such that the dynamics assures poor world utility when WL rewards are used.
This is precisely the type of problem that macrolearning is designed
to correct.

To investigate the efficacy of the macrolearning, two sets of separate
experiments were conducted.  In the first one the reward matrix $R$ is
chosen so that when leader and follower agents are placed in separate
subworlds, leaders maximizing their WL reward results in minimal world
reward. In contrast to the bar problem experiments, in these
experiments, due to the coupling between leaders and followers, having
each agent be its own subworld would mean badly violating
constraint-alignment.  (The idea is that by changing subworld
memberships macrolearning can correct this and thereby induce
constraint-alignment.)  In the second set of experiments, rather than
hand-crafting the worst-case reward matrix, we investigate the
efficacy of macrolearning for a broader spectrum of reward matrices
generated randomly.

For the worst-case reward matrix, the subworld membership that results 
in the most constraint-aligned system is when a leader 
and associated followers belong to the same subworld.  
The least constraint-alignment occurs when the leader and associated followers 
are all in separate subworlds.  
In addition to these two cases we also investigate random subworld
membership. We investigate all these cases
both with and without macrolearning, and with both worst case
and a full spectrum of random reward matrices. 
The microlearning in these experiments is the same as in the
bar problem. All leader-follower experiments use the WL subworld reward. 

When macrolearning is used, it is implemented after the microlearning
has run for a specified number of weeks. The macrolearning works by
estimating the correlations between the nights picked by the
agents. This is done by examining the attendances of the agents over
the preceding weeks.  The agents estimated to be the most correlated
with one another are grouped into the same subworld, with the
restriction that the number of agents per subworld is always three.

\begin{figure*}[ht]
   \centerline{\mbox{
      \psfig{figure={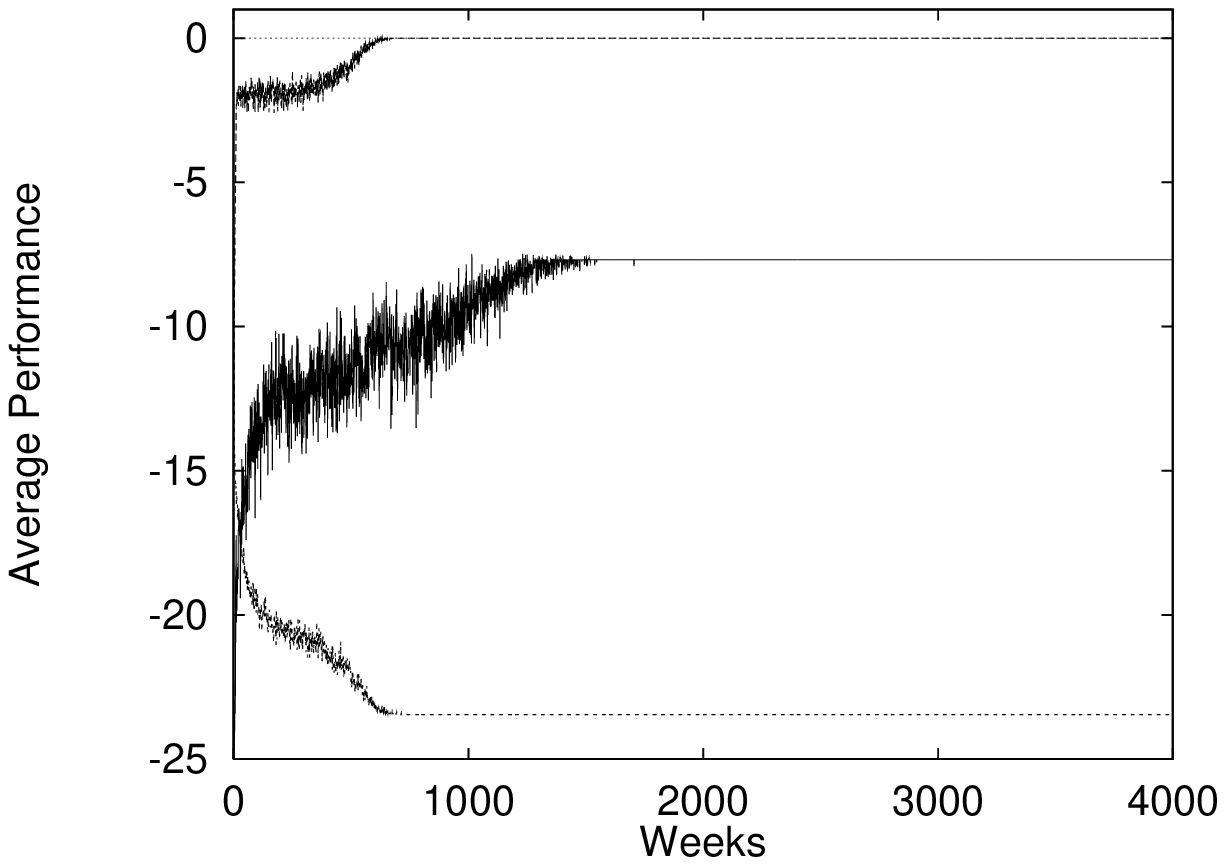},width=3.0in,height=2.25in}
      \hspace{0.1in}
      \psfig{figure={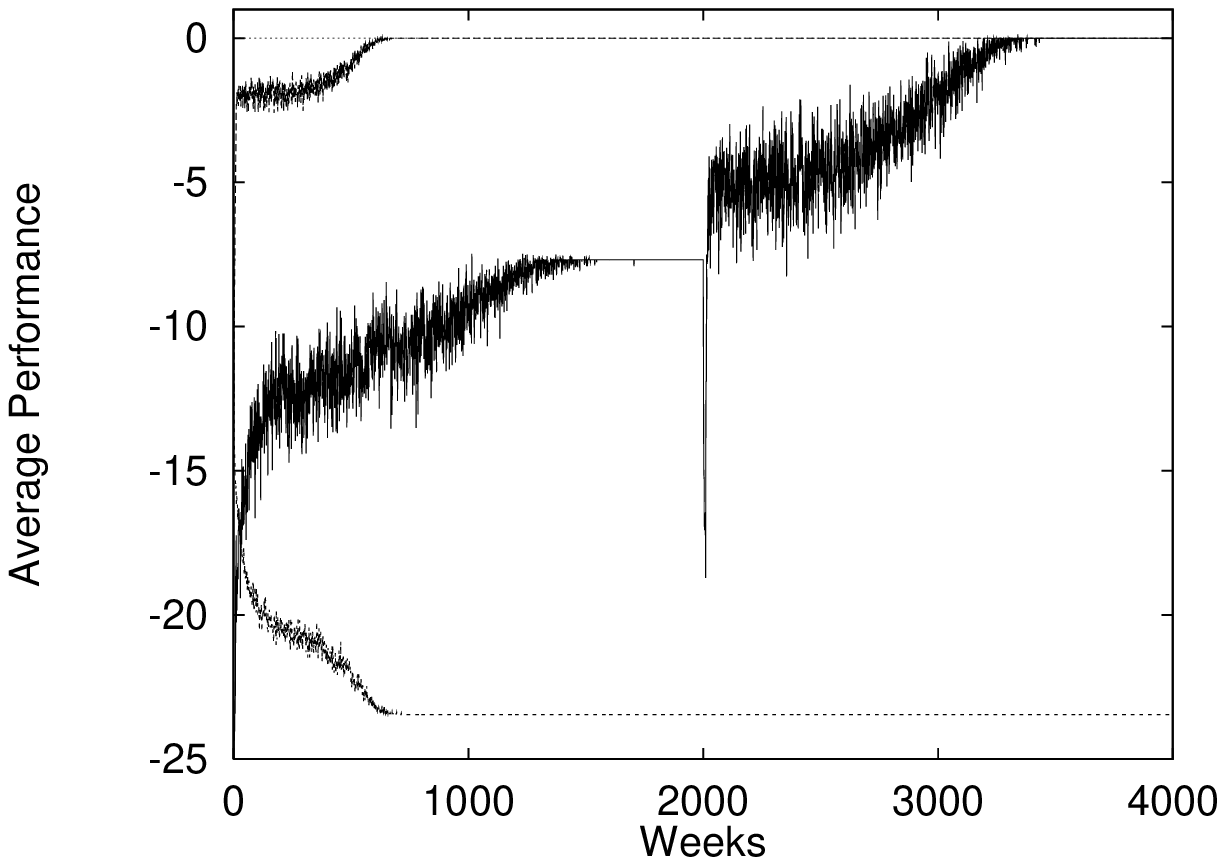},width=3.0in,height=2.25in}
   }}
   \caption{Leader-follower problem for full spectrum reward matrices. 
   The ordering of the plots is exactly as in Figure 3. 
   Macrolearning is applied at 2000 weeks, in the right plot.} 
   \label{fig:randomreward}
\end{figure*}

\subsection{Results}
Figure~\ref{fig:worstreward}  presents performance averaged over 
50 runs for world reward
as a function of weeks using the worst case reward matrix.  
For comparison purposes, the top curve represents the 
case where the leaders  and their followers 
are placed in the same subworld so that there is perfect
constraint-alignment.  The bottom curve represents the
other extreme where the leaders and their followers are placed in different
subworlds resulting in minimal constraint-alignment.  
In both plots, the middle curve shows the performance of
randomly initialized subworlds  with (right) and without (left) macrolearning.

The performance for randomly assigned
subworlds differs only slightly from that of having each agent 
be its own subworld; both start with poor values
of world reward that deteriorates with time.  
However, when macrolearning is performed on systems with initially 
random subworld
assignments, the system quickly rectifies itself and converges to
optimal performance.  This is reflected by the sudden vertical jump
through the middle of the right plot at 500 weeks, the point
at which macrolearning changes the subworld memberships.
This demonstrates that by changing the subworld memberships macrolearning 
results in perfect constraint-alignment, so that 
the WL subworld reward function quickly induces the
maximal value of the world reward. 

Figure~\ref{fig:randomreward} presents performance averaged over 
50 runs for world reward as a function of weeks using a spectrum 
of reward matrices selected at random.  
The ordering of the plots is exactly as in Figure~\ref{fig:worstreward}.
Macrolearning is applied at 2000 weeks, in the right plot.
The simulations in Figure~\ref{fig:randomreward}  were lengthened from
those in Figure~\ref{fig:worstreward} because the convergence time are
longer.

The macrolearning algorithm results in a
transient degradation in performance at 2000 weeks 
followed by convergence to the optimal.  Without macrolearning 
the system's performance no longer varies after 2000 weeks. 
Combined with the results presented in Figure 2, 
these experiments demonstrate that the COIN construction of macrolearning
works quite well.

\section{CONCLUSION}
\label{sec:conc}
For MAS's to fulfill their full potential, even when used in large
systems having strong limitations on inter-agent interactions and
communication, a way is needed to automatically configure/update the system 
to achieve the provided global goal. One
approach to this problem is to have each agent run a reinforcement
learning algorithm and then configure/update the associated reward functions (and
other characteristics) of the agents so as to facilitate achievement
of the global goal. This is the central concept embodied in COINs.

In this paper we presented a summary of (a portion of) the mathematical framework of
COINs. We then presented two sets of experiments empirically
validating the predictions of that framework. The first set of
experiments considered difficult variants of Arthur's famous El Farol
Bar problem. These experiments showed that even when the conditions
required by the theorems of how to initialize a COIN do not hold
exactly, they often hold well enough so that they can be applied with
confidence. In those experiments, the COINs quickly achieved optimal
performance despite the local nature of the information available to
each agent, and in particular the COINs automatically avoided the
tragedy of the commons that was designed into those experiments. This
was not true when the reward functions were either set in a ``naive''
manner or were all set to equal the global reward.

The second set of experiments considered leader-follower problems that
were hand-designed to cause maximal difficulty for those COIN
initialization theorems. These experiments explicitly tested the
run-time updating procedures of the COIN framework for overcoming such
initialization problems. Here, as expected, the initial performance of
the COIN was quite poor. However once the updating procedures were
brought online, performance quickly rose to optimal. Again, this was
in contrast to the case where the reward functions were either set in
a ``naive'' manner or were all set to equal the global reward.

The conclusion of these experiments is that the prescriptions of
the COIN framework for how to configure a large MAS often apply even
when the exact conditions required by the associated theorems do not
hold. Moreover, in those relatively unusual circumstances when the
initialization prescriptions of the COIN theorems do not result in
optimal global performance, the run-time updating component of the
framework can rectify the situation, so that optimal performance is
achieved.

\noindent {\bf Acknowledgments:} The authors would like to thank 
Ann Bell, Hal Duncan and Jeremy Frank for helpful discussions.

\bibliographystyle{plain}

\end{document}